\documentclass[12pt]{article}%
\usepackage{citesort}
\usepackage{amsmath}
\usepackage{graphicx}%
\usepackage{amsfonts}%
\usepackage{amssymb}
\begin{document}
\bigskip

\begin{quotation}
\end{quotation}

\vspace{1cm}

\begin{center}
{\Huge Production of gluon jets in pp collisions by double pomeron exchange }

{\Huge in the Landshoff-Nachtmann model}
\end{center}

\vskip                               3mm

\begin{center}
Adam Bzdak\footnote{e-mail: bzdak@th.if.uj.edu.pl}{\large \textbf{\textsf{ }}}
\end{center}

\vskip                                                       3mm

\begin{center}
M. Smoluchowski Institute of Physics, Jagiellonian University \newline
Reymonta 4, 30-059 Krak\'{o}w, Poland
\end{center}

\vskip                                                      .5cm

\begin{quotation}
Using the Landshoff-Nachtmann two-gluon exchange model of the pomeron, the
double pomeron exchange contribution to production of gluon pairs in the
central region of rapidity is calculated. The results are compared with those
for production of quark-antiquark pairs.
\end{quotation}

\section{Introduction}

Production of the Higgs boson by double pomeron exchange was recently studied
by several authors \cite{Ludzie}. Unfortunately the published theoretical
estimates of the cross-sections are widely different and thus do not permit to
obtain reliable predictions needed for future experiments. This reflects our
present limited understanding of the nature of the diffractive (pomeron) reactions.

\begin{figure}[h]
\begin{center}
\includegraphics[width=7cm,height=4cm]{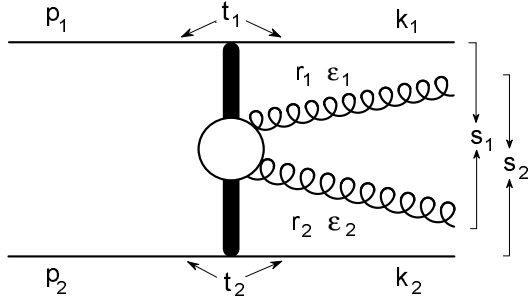}
\end{center}
\caption{Production of two gluons by double pomeron exchange.}%
\label{ogolny}%
\end{figure}

One way to reduce this ambiguity is to calculate the cross-section for other
double pomeron exchange processes and compare them with existing data. In the
present paper we study this problem using the Landshoff-Nachtmann model of the
pomeron \cite{Land-Nacht} to calculate the production of two large transverse
momentum gluon jets shown in Fig. \ref{ogolny}. In this way we hope to obtain
a better estimate of the Higgs production studied in this model some time ago
\cite{Bial-Land}.

Our calculations follow closely the method used in \cite{Bial-Szer}%
\cite{Szere} where the cross-section for production of heavy quark-antiquark
pairs was calculated. The results for two gluon jets, together with those for
quark jets, allow to obtain the full cross-section for double-diffractive jet
production in the Landshoff-Nachtmann model.

\newpage

\section{The relevant diagrams}

\begin{figure}[h]
\begin{center}
\includegraphics[width=3.80cm,height=2cm]{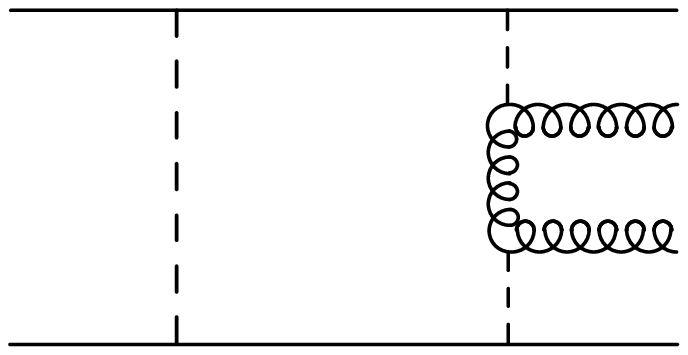}  \hspace{0.3cm}
\includegraphics[width=3.80cm,height=2cm]{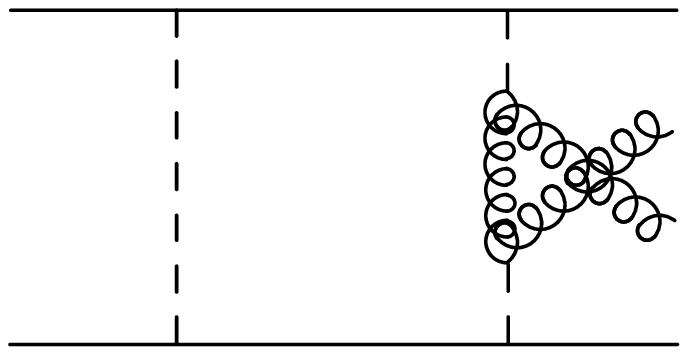}  \hspace{0.3cm}
\includegraphics[width=3.80cm,height=2cm]{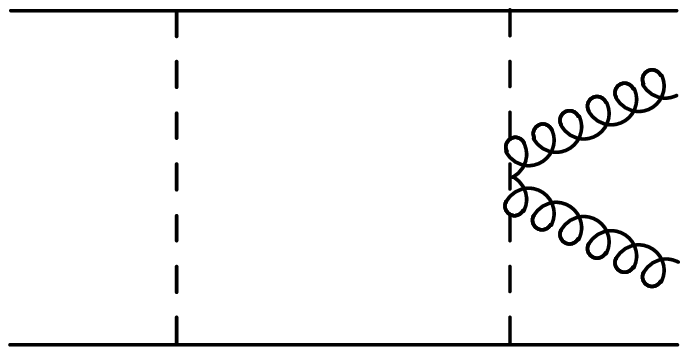}
\par
\vspace{0.5cm} \includegraphics[width=3.80cm,height=2cm]{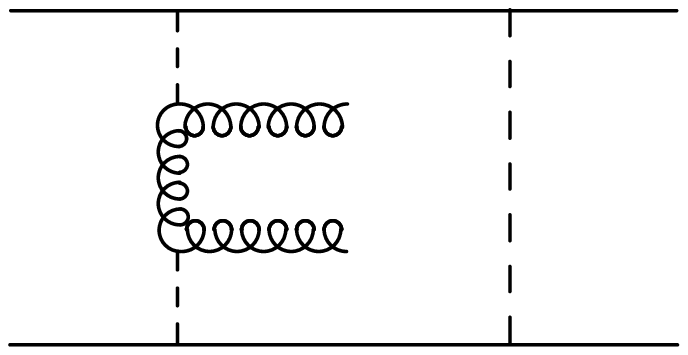}
\hspace{0.3cm} \includegraphics[width=3.80cm,height=2cm]{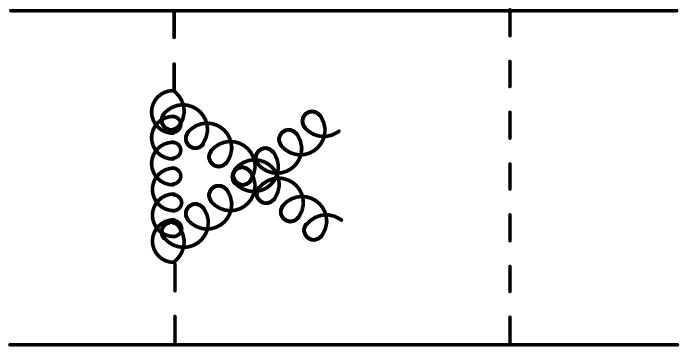}
\hspace{0.3cm} \includegraphics[width=3.80cm,height=2cm]{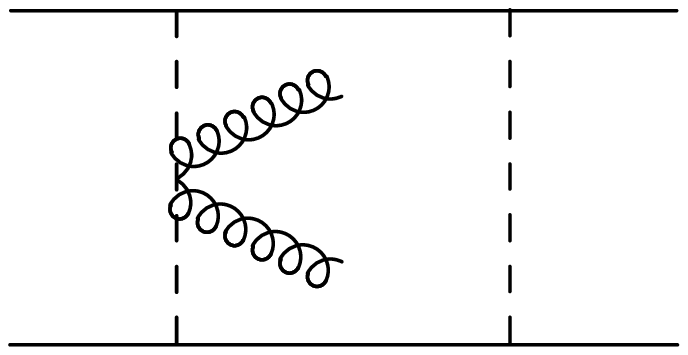}
\par
\vspace{0.5cm} \includegraphics[width=3.80cm,height=2cm]{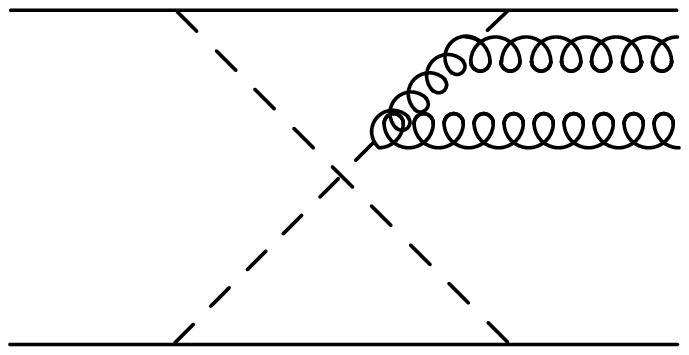}
\hspace{0.3cm} \includegraphics[width=3.80cm,height=2cm]{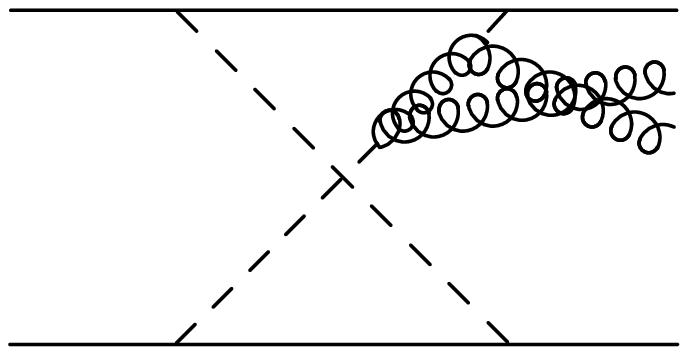}
\hspace{0.3cm} \includegraphics[width=3.80cm,height=2cm]{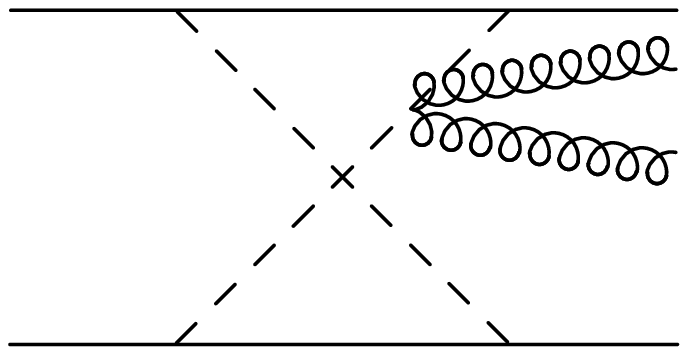}
\par
\vspace{0.5cm} \includegraphics[width=3.80cm,height=2cm]{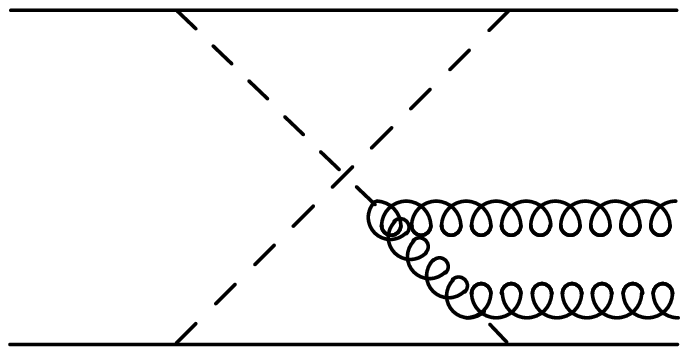}
\hspace{0.3cm} \includegraphics[width=3.80cm,height=2cm]{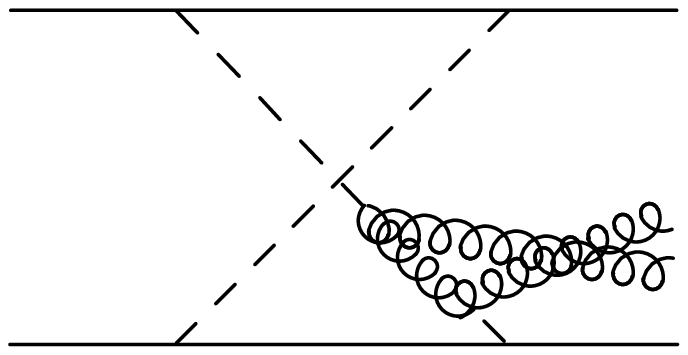}
\hspace{0.3cm} \includegraphics[width=3.80cm,height=2cm]{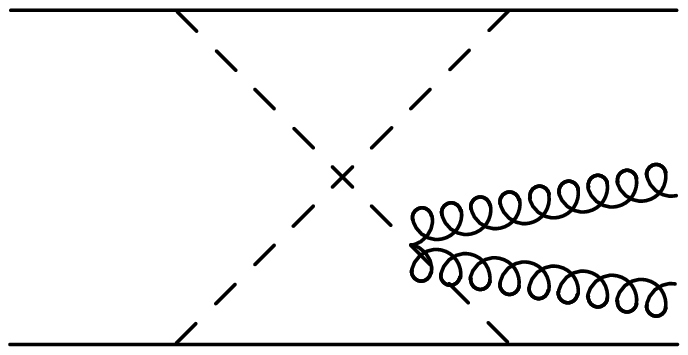}
\end{center}
\caption{The diagrams which have to be taken into account in the process of
the gluon pair production by double pomeron exchange. The dashed lines
represent the exchange of the nonperturbative gluons.}%
\label{12}%
\end{figure}

In the Landshoff-Nachtmann model \cite{Land-Nacht} the pomeron is described as
an exchange of two ''nonperturbative'' gluons, which takes place between a
pair of quarks of colliding hadrons. Thus to calculate the cross-section for
reaction $pp\longrightarrow$ $pp$ $+$ $gg$ by double pomeron exchange we have
to take into account twelve diagrams presented in Fig. \ref{12}.

\vspace{0.5cm}

\begin{figure}[h]
\begin{center}
\includegraphics[width=3.80cm,height=2cm]{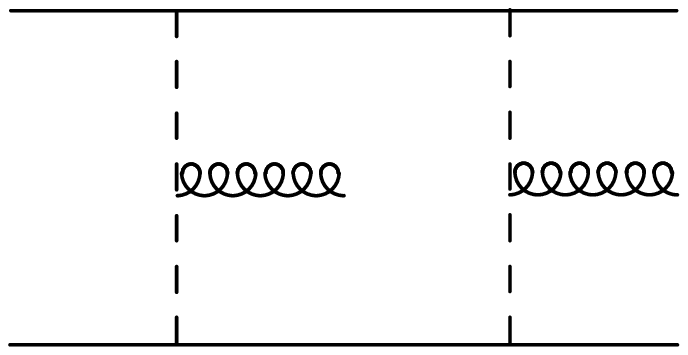}  \hspace{0.5cm}
\includegraphics[width=3.80cm,height=2cm]{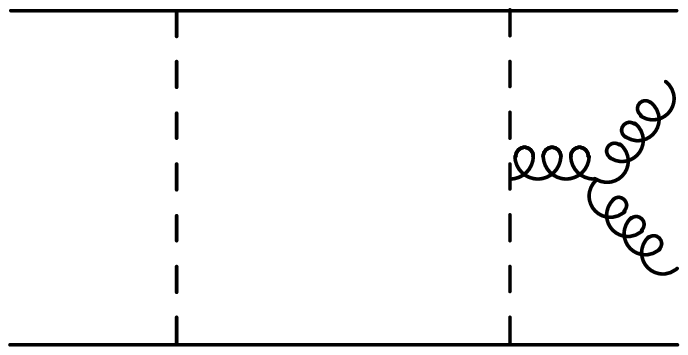}
\end{center}
\par
\vspace{0cm} \caption{The examples of the diagrams which do not contribute.}%
\label{nie}%
\end{figure}

Other diagrams, like these presented in Fig. \ref{nie}, either do not
contribute significantly because of large transverse momentum on one of the
quark internal lines or simply vanish because pomeron is the colour singlet.

\vspace{0.6cm}

Following \cite{Bial-Land} we assume that up to highest powers of $s/s_{1}$
and $s/s_{2},$ the total amplitude given by diagrams in Fig. \ref{12} is of
the form:%

\begin{equation}
M_{fi}=M_{fi}|_{t_{1}=t_{2}=0}\left(  \frac{s}{s_{1}}\right)  ^{\alpha_{2}%
-1}\left(  \frac{s}{s_{2}}\right)  ^{\alpha_{1}-1}. \label{Mfi}%
\end{equation}
Here $\alpha_{1}=\alpha\left(  t_{1}\right)  ,$ $\alpha_{2}=\alpha\left(
t_{2}\right)  $ and $\alpha\left(  t\right)  =1+\epsilon+\alpha^{\prime}t$ is
the pomeron Regge trajectory with $\epsilon\approx0.08,$ $\alpha^{\prime
}=0.25$ GeV$^{-2}$ ( $s_{1},$ $s_{2},$ $t_{1},$ $t_{2}$ are marked in Fig.
\ref{ogolny}). So, in order to calculate the amplitude for production of two
gluons by double pomeron exchange, it is enough to take diagrams from Fig.
\ref{12} at $t_{1}=t_{2}=0.$

Even with this simplification, however, the direct evaluation of all diagrams
of Fig. \ref{12} is still tedious. Fortunately, as shown in \cite{Zakrzewski}
and exploited in \cite{Bial-Land}, it turns out that up to terms of the order
$\epsilon$, the matrix element $M_{fi}|_{t_{1}=t_{2}=0}$ is equal to the sum
of the following three diagrams:

\vspace{0.6cm}

\begin{figure}[h]
\begin{center}
\includegraphics[width=4.80cm,height=3.5cm]{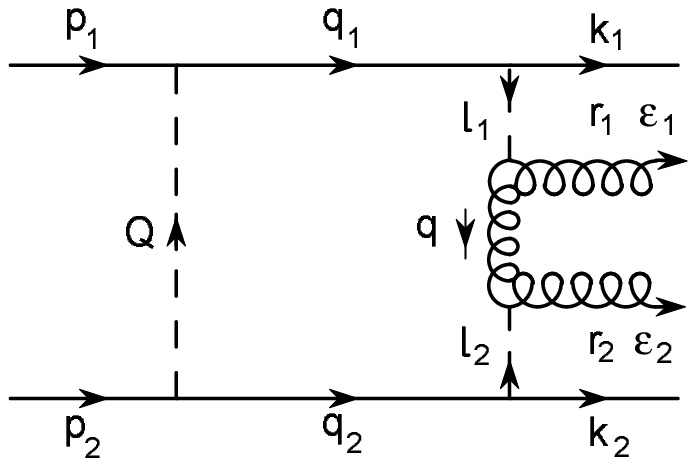}  \hspace{0.3cm}
\includegraphics[width=4.80cm,height=3.5cm]{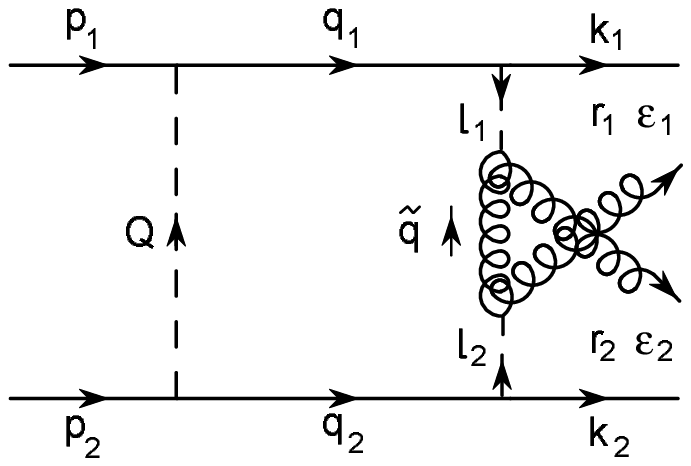}  \hspace{0.3cm}
\includegraphics[width=4.80cm,height=3.5cm]{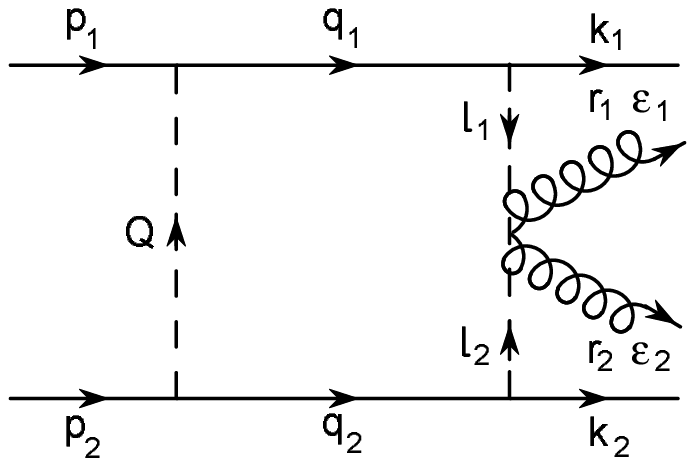}
\end{center}
\caption{Putting the quarks lines $q_{1}$ and $q_{2}$ on shell is equivalent
to calculating the amplitude for production of gluon pair by double pomeron
exchange at $t_{1}=t_{2}=0.$ }%
\label{ok}%
\end{figure}

\noindent where the lines $q_{1}$ and $q_{2}$ are put on shell and
$t_{1}=t_{2}=0.$

The calculation is performed in two steps. First, we calculate $M_{fi}%
|_{t_{1}=t_{2}=0}$ for colliding quarks, given by the sum of the diagrams
presented in Fig. \ref{ok} and use (\ref{Mfi}) to introduce the Regge
behavior. To obtain the cross-section the presence of three quarks in the
proton must be taken into account.

\section{Matrix element}

We take the masses of the light quarks to be approximately zero and, following
\cite{Bial-Land}, use the Sudakov parametrization for momenta shown in Fig.
\ref{ok}:%

\begin{align}
Q  &  =\frac{x}{s}p_{1}+\frac{y}{s}p_{2}+v,\nonumber\\
k_{1}  &  =x_{1}p_{1}+\frac{y_{1}}{s}p_{2}+v_{1},\nonumber\\
k_{2}  &  =\frac{x_{2}}{s}p_{1}+y_{2}p_{2}+v_{2},\nonumber\\
r_{2}  &  =x_{g}p_{1}+y_{g}p_{2}+v_{g}, \label{parametr}%
\end{align}
where $v,$ $v_{1},$ $v_{2},$ $v_{g}$ are two-dimensional four-vectors.

Let us express $M_{fi}|_{t_{1}=t_{2}=0}$ as the following sum:%

\begin{equation}
M_{fi}|_{t_{1}=t_{2}=0}\equiv M_{fi}^{\left(  3\right)  }+M_{fi}^{\left(
4\right)  }, \label{rozklad}%
\end{equation}
where $M_{fi}^{\left(  3\right)  }$ is the sum of the first two diagrams
presented in Fig. \ref{ok}, and $M_{fi}^{\left(  4\right)  }$ being last
diagram in Fig. \ref{ok}. Following closely the method used in \cite{Szere} we
obtain (see appendix):%

\begin{equation}
M_{fi}^{\left(  4\right)  }=-A\frac{4}{12}\delta_{s,c}^{if}\delta_{uw}\pi\tau
s\varepsilon_{1}\varepsilon_{2}, \label{Mfi4-pocza}%
\end{equation}%

\begin{equation}
M_{fi}^{\left(  3\right)  }=-A\frac{4}{12}\delta_{s,c}^{if}\delta_{uw}\int
d^{2}\vec{v}f\left(  \vec{v}\right)  \exp\left(  \frac{-\vec{v}^{2}}{\tau
}\right)  , \label{Mfi3-pocza}%
\end{equation}
where:%

\begin{align}
f\left(  \vec{v}\right)   &  \equiv\frac{1}{q^{2}}\left(  s^{2}\varepsilon
_{1}\varepsilon_{2}\left(  \delta_{2}-y_{g}\right)  x_{g}-2s\varepsilon
_{1}v\varepsilon_{2}v+2s\varepsilon_{1}v\varepsilon_{2}p_{2}\left(  \delta
_{2}-y_{g}\right)  -2s\varepsilon_{2}v\varepsilon_{1}p_{2}\left(  \delta
_{2}-y_{g}\right)  \right) \nonumber\\
&  +\frac{1}{\tilde{q}^{2}}\left(  s^{2}\varepsilon_{1}\varepsilon_{2}\left(
\delta_{1}-x_{g}\right)  y_{g}-2s\varepsilon_{1}v\varepsilon_{2}%
v+2s\varepsilon_{2}v\varepsilon_{1}p_{2}y_{g}-2s\varepsilon_{1}v\varepsilon
_{2}p_{2}y_{g}\right)  . \label{f(v)}%
\end{align}
Here $A=i(D_{0}G^{2})^{3}g^{2}/(2\pi)^{2}G^{2}$, $G$ and $g$ are the
non-perturbative and perturbative quark-gluon couplings, $\tau=\mu^{2}/\left(
1+x_{1}+y_{2}\right)  $. $D\left(  p^{2}\right)  =D_{0}\exp\left(  p^{2}%
/\mu^{2}\right)  $ with $\mu\approx1$ GeV and $G^{2}D_{0}\approx30$
GeV$^{-1}\mu^{-1},$ is the non-perturbative gluon propagator. By $\delta_{1}$
and $\delta_{2}$ we denote $1-x_{1}$ and $1-y_{2}$ respectively. The indices
$u$ and $w$ denote the colour of the produced gluons. $\varepsilon_{1}$ and
$\varepsilon_{2}$ are the polarization vectors of gluons chosen to satisfy the relations:%

\begin{equation}
\sum_{\lambda=1,2}\varepsilon_{i}^{\mu}\left(  \lambda\right)  \varepsilon
_{i}^{\nu}\left(  \lambda\right)  =-g^{\mu\nu}+\frac{p_{1}^{\mu}r_{i}^{\nu
}+p_{1}^{\nu}r_{i}^{\mu}}{p_{1}r_{i}}\qquad i=1,2. \label{polariz}%
\end{equation}
So that, in addition to the constraints $\varepsilon_{1}r_{1}=\varepsilon
_{2}r_{2}=0$, we have further constraints: $\varepsilon_{1}p_{1}%
=\varepsilon_{2}p_{1}=0$.

Performing the integration in (\ref{Mfi3-pocza}) is a rather difficult task.
Following \cite{Szere} we expand the integral over $d^{2}\vec{v}$ in power series:%

\begin{equation}
\int d^{2}\vec{v}f\left(  \vec{v}\right)  \exp\left(  \frac{-\vec{v}^{2}}%
{\tau}\right)  =a\pi\tau+\sum_{j}\frac{\pi\tau^{2}}{2}a_{jj}+...,
\label{rozwi}%
\end{equation}
where $a$ and $a_{jj}$ are coefficients of expansion of $f\left(  \vec
{v}\right)  $:%

\begin{equation}
a=f\left(  \vec{v}=0\right)  ,\qquad a_{jj}=\frac{1}{2}\frac{\partial
^{2}f\left(  \vec{v}\right)  }{\partial\vec{v}^{j}\partial\vec{v}^{j}}%
|_{\vec{v}=0}. \label{coeffic}%
\end{equation}
One can check that in the first term of (\ref{rozwi}) $M_{fi}^{\left(
3\right)  }$ and $M_{fi}^{\left(  4\right)  }$ exactly cancel each other. So
we are forced to calculate the second term. After rather lengthy calculations
we obtain:%

\begin{equation}
M_{fi}|_{t_{1}=t_{2}=0}=-A\frac{4}{12}\delta_{s,c}^{if}\delta_{uw}\pi\tau
^{2}\left\{  -\frac{1}{x_{g}y_{g}}\varepsilon_{1}\varepsilon_{2}%
+\frac{2}{sx_{g}y_{g}}\frac{\delta_{2}}{\delta_{1}}\varepsilon_{1}%
p_{2}\varepsilon_{2}p_{2}\right\}  . \label{am1}%
\end{equation}

Taking the square of the amplitude (\ref{am1}), averaging and summing over
spins, colours and polarizations one arrives at the simple formula:%

\[
\overline{\left|  M_{fi|}{}_{t_{1}=t_{2}=0}\right|  ^{2}}=\frac{H}{x_{g}%
^{2}y_{g}^{2}},
\]
where:%

\begin{equation}
H=\frac{8}{(12)^{2}}2\left(  \frac{4\pi\left(  D_{0}G^{2}\right)  ^{3}\mu^{4}%
}{9\left(  2\pi\right)  ^{2}}\right)  ^{2}\left(  \frac{g^{2}}{G^{2}}\right)
^{2}. \label{H}%
\end{equation}
Factors $8$ and $\left(  \frac{1}{12}\right)  ^{2}$ are related to the summing
over colours of produced gluons and to the square of colour factor of diagrams
represented by $M_{fi}^{\left(  3\right)  },$ respectively.

Taking into account (\ref{Mfi}) where $(s/s_{1})=(1/\delta_{2})$,
$(s/s_{2})=\left(  1/\delta_{1}\right)  $, $t_{1}=-\vec{v}_{1}^{2}$ and
$t_{2}=-\vec{v}_{2}^{2}$ we obtain:%

\begin{equation}
\overline{\left|  M_{fi}\right|  ^{2}}=\frac{H}{\left(  x_{g}y_{g}\right)
^{2}\left(  \delta_{1}\delta_{2}\right)  ^{2\epsilon}\delta_{1}^{2\alpha
^{\prime}t_{1}}\delta_{2}^{2\alpha^{\prime}t_{2}}}\exp\left(  2\beta\left(
t_{1}+t_{2}\right)  \right)  . \label{Mfi-koniec}%
\end{equation}
The factor $\exp\left(  2\beta\left(  t_{1}+t_{2}\right)  \right)  $ takes
into account the effect of the momentum transfer dependence of the
non-perturbative gluon propagator with $\beta$ $=$ $1$ GeV$^{-2}$
\cite{Factor}\cite{Bial-Janik}.

\section{Cross-section}

Having calculated $\overline{\left|  M_{fi}\right|  ^{2}}$ we can write the
formula for the cross-section:%

\begin{equation}
\sigma=\frac{81}{2s\left(  2\pi\right)  ^{8}2!}\int\overline{|M_{fi}|^{2}%
}\left[  F\left(  t_{1}\right)  F\left(  t_{2}\right)  \right]  ^{2}dPH,
\label{cro-sec-ogolny}%
\end{equation}
where the factor $81$ takes into account the presence of three quarks in each
proton, $2!$ is an identical final state particle phase space
factor\footnote{We would like to thank J. R. Cudell for a correspondence about
this point.} and $F\left(  t\right)  $ is the nucleon formfactor approximated by:%

\begin{equation}
F\left(  t\right)  =\exp\left(  \lambda t\right)  \label{formfactor}%
\end{equation}
with $\lambda=$ $2$ GeV$^{-2}$. Differential phase-space factor $dPH$ has the form:%

\begin{align}
dPH  &  =d^{4}k_{1}\delta\left(  k_{1}^{2}\right)  d^{4}k_{2}\delta\left(
k_{2}^{2}\right)  d^{4}r_{1}\delta\left(  r_{1}^{2}\right)  d^{4}r_{2}%
\delta\left(  r_{2}^{2}\right)  \times\nonumber\\
&  \Theta\left(  k_{1}^{0}\right)  \Theta\left(  k_{2}^{0}\right)
\Theta\left(  r_{1}^{0}\right)  \Theta\left(  r_{2}^{0}\right)  \delta
^{(4)}\left(  p_{1}+p_{2}-k_{1}-k_{2}-r_{1}-r_{2}\right)  . \label{fazowa}%
\end{align}

It turns out that a substantial part of the integrations can be performed
analytically \cite{Szere}. Denoting by $E_{\min}^{\intercal}$ the minimum
value of transverse energy of the produced gluon (we integrate over
$E^{\intercal}\geq E_{\min}^{\intercal}\neq0$) and by $\Delta$ the maximum
value of the energy loss of the initial hadrons $(\delta_{1,2}<\Delta
\sim0.1),$ we obtain:
\begin{align}
\sigma\left(  E_{\min}^{\intercal}\right)   &  =C\overset{h\left(
\Delta\right)  }{\underset{0}{%
{\displaystyle\int}
}}dxx\left(  1-x^{2}\right)  ^{2\epsilon}\left(  \ln\frac{1+x}{1-x}%
+\frac{2x}{1-x^{2}}\right) \nonumber\\
&  \frac{1}{\frac{\lambda+\beta}{\alpha^{\prime}}+\frac{1}{2}\ln\frac{1-x^{2}%
}{\delta^{2}}}\ln\frac{\frac{\lambda+\beta}{\alpha^{\prime}}+\ln
\frac{\Delta\left(  1-x^{2}\right)  }{\delta^{2}}}{\frac{\lambda+\beta}%
{\alpha^{\prime}}+\ln\frac{1}{\Delta}}, \label{cro-sec-gg}%
\end{align}
where $h\left(  \Delta\right)  =\sqrt{1-\delta^{2}/\Delta^{2}},$ $\delta
^{2}=4(E_{\min}^{\intercal})^{2}/s$ and:%

\[
C=9\pi\left(  \frac{s}{4(E_{\min}^{\intercal})^{2}}\right)  ^{2\epsilon
}\left(  \frac{\pi^{2}\left(  D_{0}G^{2}\right)  ^{3}\mu^{4}}{18\left(
2\pi\right)  ^{6}E_{\min}^{\intercal}\alpha^{\prime}}\right)  ^{2}\left(
\frac{g^{2}}{G^{2}}\right)  ^{2}.
\]

The major uncertainty in above result is the value of the non-perturbative
coupling constant $G.$

Finally, let us remind the result for quark-antiquark production by double
pomeron exchange \cite{Bial-Szer}:%

\begin{align}
\sigma\left(  (k_{\min}^{\intercal})^{2}\right)   &  =\widetilde{C}%
\overset{\widetilde{h}\left(  \Delta\right)  }{\underset{0}{%
{\displaystyle\int}
}}dxx\left(  1-m^{2}\frac{1-x^{2}}{m^{2}+(k_{\min}^{\intercal})^{2}}\right)
\left(  1-x^{2}\right)  ^{1+2\epsilon}\left(  \ln\frac{1+x}{1-x}%
+\frac{2x}{1-x^{2}}\right) \nonumber\\
&  \frac{1}{\frac{\lambda+\beta}{\alpha^{\prime}}+\frac{1}{2}\ln\frac{1-x^{2}%
}{\widetilde{\delta}^{2}}}\ln\frac{\frac{\lambda+\beta}{\alpha^{\prime}}%
+\ln\frac{\Delta\left(  1-x^{2}\right)  }{\widetilde{\delta}^{2}}%
}{\frac{\lambda+\beta}{\alpha^{\prime}}+\ln\frac{1}{\Delta}},
\label{cro-sec-qq}%
\end{align}
where $\widetilde{h}\left(  \Delta\right)  =\sqrt{1-\widetilde{\delta}%
^{2}/\Delta^{2}},$ $\widetilde{\delta}^{2}=4\left(  m^{2}+(k_{\min}%
^{\intercal})^{2}\right)  /s$, $m$ is the mass of the produced quark, and:%

\[
\widetilde{C}=\frac{1}{3}\pi\left(  \frac{s}{4\left(  m^{2}+(k_{\min
}^{\intercal})^{2}\right)  }\right)  ^{2\epsilon}\left(  \frac{\pi^{2}\left(
D_{0}G^{2}\right)  ^{3}\mu^{4}m}{18\left(  2\pi\right)  ^{6}\alpha^{\prime
}\left(  m^{2}+(k_{\min}^{\intercal})^{2}\right)  }\right)  ^{2}\left(
\frac{g^{2}}{G^{2}}\right)  ^{2}.
\]
By $k_{\min}^{\intercal}$ we denote the minimum value of transverse momentum
of the produced quark.

\section{Numerical results and discussion}

Let us remind the following numbers that appear in (\ref{cro-sec-gg}) and
(\ref{cro-sec-qq}): $\epsilon=0.08,$ $\alpha^{\prime}=0.25$ GeV$^{-2},$
$\mu=1$ GeV, $G^{2}D_{0}=30$ GeV$^{-1}\mu^{-1},$ $\lambda=$ $2$ GeV$^{-2},$
$\beta$ $=$ $1$ GeV$^{-2},$ $\Delta=0.1.$ The minimum value of transverse
energy of the produced gluon $E_{\min}^{\intercal}$ is chosen at $5$ $,7$ and
$10$ GeV . At Tevatron energy, $\sqrt{s}=1.8$ TeV, the results for the
cross-section $\sigma$ (multiplied by $\left(  G^{2}/4\pi\right)  ^{2}$) are
as follows:

\vspace{0.5cm}

\begin{table}[h]
\begin{center}%
\begin{tabular}
[c]{|c|c|c|c|}\hline\hline
$\sqrt{s}$ $[$TeV$]$ & $E_{\min}^{\intercal}=5$ GeV & $E_{\min}^{\intercal}=7$
GeV & $E_{\min}^{\intercal}=10$ GeV\\\hline
$1.8$ & $\left(  G^{2}/4\pi\right)  ^{2}\sigma=3$ $\mu$b & $\left(  G^{2}%
/4\pi\right)  ^{2}\sigma=0.9$ $\mu$b & $\left(  G^{2}/4\pi\right)  ^{2}%
\sigma=0.3$ $\mu$b\\\hline
\end{tabular}
\end{center}
\caption{The numerical results for the cross-section for production of gluon
pairs by double pomeron exchange. The calculation is performed at Tevatron
energy $\sqrt{s}=1.8$ TeV. }%
\label{ggTevatron}%
\end{table}

The running coupling constant $g^{2}/4\pi$ was evaluated at $2E_{\min
}^{\intercal},$ \emph{i.e. }$0.17,$ $0.15$ and $0.14$ for $E_{\min}%
^{\intercal}=5,7,10$ GeV respectively.

\begin{figure}[h]
\begin{center}
\includegraphics[width=9cm,height=7cm]{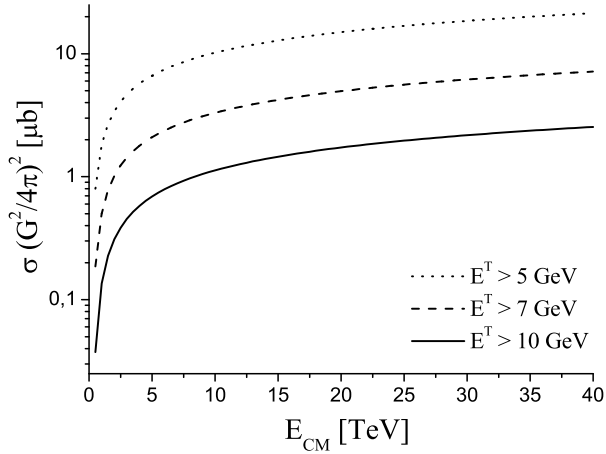}
\end{center}
\caption{Double pomeron exchange contribution to the cross-section for the
gluon pairs production process versus the center mass energy.}%
\label{ggwyk}%
\end{figure}

The dependence of the cross-section versus the center mass energy
$E_{CM}=\sqrt{s}$ for the range of $\sqrt{s}$ taken from $0.5$ TeV to $40$ TeV
is shown in Fig. \ref{ggwyk}.

\vspace{0.5cm}

In order to obtain the total cross-section for production of jets by double
pomeron exchange, we have to consider the quark-antiquark production process.
It is obvious that for the case in which the minimum value of transverse
momentum of the produced quark $k_{\min}^{\intercal}$ $=0,$ the relations
hold: $\sigma_{u\bar{u}}\simeq\sigma_{d\bar{d}}>\sigma_{s\bar{s}}%
>>\sigma_{c\bar{c}}>$ $\sigma_{b\bar{b}}>$ $\sigma_{t\bar{t}}.$ The values for
$\sigma_{c\bar{c}},$ $\sigma_{b\bar{b}}$ and $\sigma_{t\bar{t}}$ are
determined by (\ref{cro-sec-qq}) (this expression is reliable for heavy quarks
production process). For more details see \cite{Bial-Szer}\cite{Szere}.

In the experiment we observe jets with some minimal $E_{\min}^{\intercal}$
$\sim$ GeV, however, and these relations are not valid. At Tevatron energy
$\sqrt{s}=1.8$ TeV and $E_{\min}^{\intercal}$ $=7$ GeV we have:

\vspace{0.5cm}

\begin{table}[h]
\begin{center}%
\begin{tabular}
[c]{|c|c|c|c|}\hline\hline
$\sqrt{s}$ $[$TeV$]$ & $\left(  G^{2}/4\pi\right)  ^{2}\sigma_{c\bar{c}}$
$\left[  \text{nb}\right]  $ & $\left(  G^{2}/4\pi\right)  ^{2}\sigma
_{b\bar{b}}$ $\left[  \text{nb}\right]  $ & $\left(  G^{2}/4\pi\right)
^{2}\sigma_{t\bar{t}}$ $\left[  \text{nb}\right]  $\\\hline
$1.8$ & $0.3$ & $3.1$ & $0$\\\hline
\end{tabular}
\end{center}
\caption{The numerical results for the cross-section for production of heavy
quark pairs by double pomeron exchange. The calculation is performed at
Tevatron energy $\sqrt{s}=1.8$ TeV and $E_{\min}^{\intercal}=7$ GeV. }%
\label{qqTevatron}%
\end{table}

As the masses of the heavy quarks we take $m_{c}=1.2$ GeV, $m_{b}=4.2$ GeV,
$m_{t}=175$ GeV (there is not enough energy to produce a $t\bar{t}$ pair). The
running coupling constant $g^{2}/4\pi$ was evaluated at $2E_{\min}^{\intercal
},$ \emph{i.e. }$0.15,$ $0.15$ and $0.1$ for $c\bar{c},$ $b\bar{b}$ and
$t\bar{t}$ respectively.

\begin{figure}[h]
\begin{center}
\includegraphics[width=9cm,height=7cm]{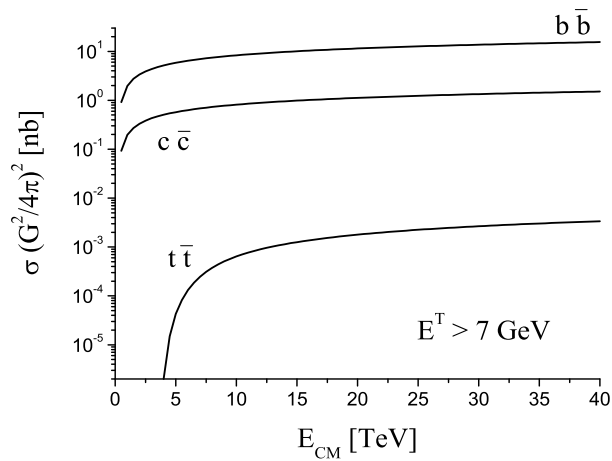}
\end{center}
\caption{Double pomeron exchange contribution to the cross-section for
production of heavy quark pairs process versus the center mass energy.}%
\label{qqwyk}%
\end{figure}

The dependence of the cross-section on the center of mass energy $E_{CM}%
=\sqrt{s}$ for the range of $\sqrt{s}$ taken from $0.5$ TeV to $40$ TeV is
shown in Fig. \ref{qqwyk}.

\vspace{0.5cm}

As we see $\sigma_{b\bar{b}}>\sigma_{c\bar{c}}.$ This is a consequence of the
fact that we observe the jets with minimal $E_{\min}^{\intercal}=7$ GeV .
Since there are very few light quarks with $E_{\min}^{\intercal}$ $\thicksim$
GeV, the cross-section for production $u\bar{u}$ $d\bar{d}$ $s\bar{s}$ pairs
with such minimal energy is negligible. To summarize, the cross-section for
quark-antiquark production with minimal $E_{\min}^{\intercal}\thicksim$ GeV is
fully dominated by the $c\bar{c}$ and $b\bar{b}$ pairs production.

One sees that $q\bar{q}$ contribution to production of jets is about three
orders of magnitude smaller than that of $gg$ contribution. Similar
observation was first made in Ref. \cite{Berera}.

Our cross sections are calculated for the case where both initial protons are
scattered quasi-elastically. If one allows for diffractive dissociation of the
incident hadrons \emph{i.e.} if one puts $\lambda=0$ in (\ref{formfactor}) the
cross sections increase about half order of magnitude.

\section{Conclusions}

We have calculated the cross-section for gluon pair production by double
pomeron exchange in the Landshoff-Nachtmann model. The obtained results were
compared with those for production of quark-antiquark pairs calculated in the
same model. It turns out that cross-section for jet production is strongly
dominated by $gg$ production.

\vspace{1cm}

I would like to thank Professor Andrzej Bia\l as for suggesting this
investigation and helpful discussions. Discussions with Prof. Maciej A. Nowak,
Prof. Jacek Wosiek, Prof. Micha\l\ Prasza\l owicz and dr Leszek Motyka are
also highly appreciated.

\section{Appendix}

In the present appendix some main steps leading to the expressions
(\ref{Mfi4-pocza}) and (\ref{Mfi3-pocza}) are more explicitly shown.

Let us denote the colour structure of the diagrams represented by
$M_{fi}^{\left(  3\right)  }$ as follows:

\begin{figure}[h]
\begin{center}
\includegraphics[width=3.90cm,height=3cm]{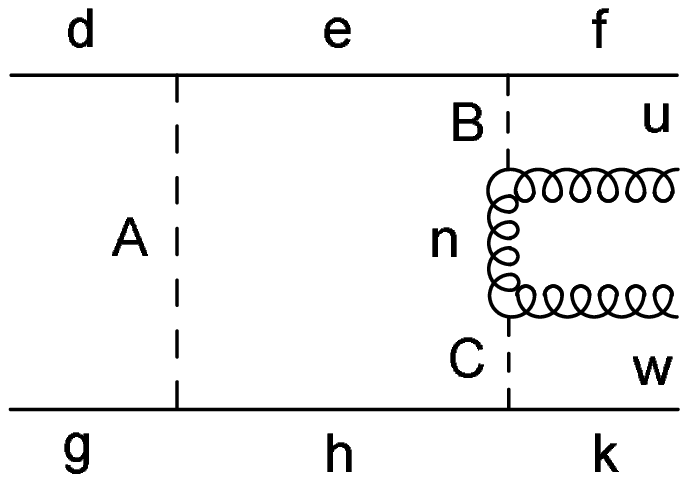}  \hspace{0.5cm}
\includegraphics[width=3.90cm,height=3cm]{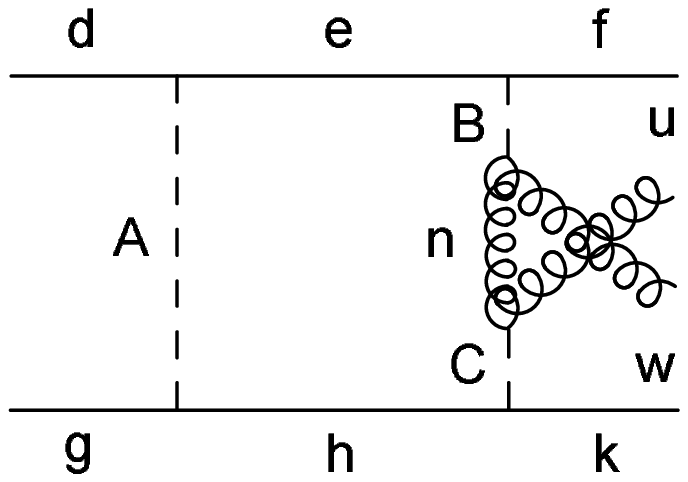}
\end{center}
\par
\vspace{0cm} \caption{The capital letters denote colour indicies of the
non-perturbative gluons and small letters colour indicies of quarks and
perturbative gluons.}%
\label{color1-2}%
\end{figure}

Performing colour calculations for the first diagram in Fig. \ref{color1-2}
(one has to remember that pomeron does not change colour of colliding quarks)
we have:%

\begin{equation}
\frac{1}{4}\left(  \lambda_{fe}^{B}\lambda_{ed}^{A}\right)  _{\text{singlet}%
}\frac{1}{4}\left(  \lambda_{kh}^{C}\lambda_{hg}^{A}\right)  _{\text{singlet}%
}f_{Bun}f_{nwC}=-\frac{1}{12}\delta_{fd}\delta_{kg}\delta_{uw},
\end{equation}
what is a consequence of:%

\begin{align}
\left(  \lambda_{fe}^{B}\lambda_{ed}^{A}\right)  _{\text{singlet}}  &
=\frac{2}{3}\delta^{AB}\delta_{fd}\nonumber\\
\left(  \lambda_{kh}^{C}\lambda_{hg}^{A}\right)  _{\text{singlet}}  &
=\frac{2}{3}\delta^{CA}\delta_{kg}. \label{singlet}%
\end{align}
One can check that for the second diagram the colour factor is the same.

The matrix element $M_{fi}^{\left(  3\right)  }$ is given by (see Fig.
\ref{ok}):%

\begin{align}
M_{fi}^{\left(  3\right)  }  &  =-\frac{A}{12}\delta_{c}^{if}\delta_{uw}%
{\displaystyle\int}
d^{4}Q\delta\left(  q_{1}^{2}\right)  \delta\left(  q_{2}^{2}\right)  D\left(
Q^{2}\right)  D\left(  l_{1}^{2}\right)  D\left(  l_{2}^{2}\right) \nonumber\\
&  \bar{u}\left(  k_{1}\right)  \gamma^{\beta}\not q  _{1}\gamma^{\nu}u\left(
p_{1}\right)  \bar{u}\left(  k_{2}\right)  \gamma^{\lambda}\not q  _{2}%
\gamma_{\nu}u\left(  p_{2}\right) \nonumber\\
&  \left\{  V_{\beta\xi\delta}\left(  l_{1},-r_{1},-q\right)  \frac{-g^{\delta
\rho}}{q^{2}}V_{\rho\sigma\lambda}\left(  q,-r_{2},l_{2}\right)
\varepsilon_{1}^{\xi}\varepsilon_{2}^{\sigma}\right. \nonumber\\
&  \left.  +V_{\beta\xi\delta}\left(  l_{1},-r_{2},\tilde{q}\right)
\frac{-g^{\delta\rho}}{\tilde{q}^{2}}V_{\rho\sigma\lambda}\left(  -\tilde
{q},-r_{1},l_{2}\right)  \varepsilon_{1}^{\sigma}\varepsilon_{2}^{\xi
}\right\}  , \label{Mfi3appen}%
\end{align}
where:%

\begin{equation}
V_{\mu_{1}\mu_{2}\mu_{3}}\left(  k_{1},k_{2},k_{3}\right)  =g_{\mu_{1}\mu_{2}%
}\left(  k_{1}-k_{2}\right)  _{\mu_{3}}+g_{\mu_{2}\mu_{3}}\left(  k_{2}%
-k_{3}\right)  _{\mu_{1}}+g_{\mu_{3}\mu_{1}}\left(  k_{3}-k_{1}\right)
_{\mu_{2}}. \label{V}%
\end{equation}
Here $A=iD_{0}^{3}G^{4}g^{2}/(2\pi)^{2}$, $G$ and $g$ are the non-perturbative
and perturbative quark-gluon couplings, $D\left(  p^{2}\right)  =D_{0}%
\exp\left(  p^{2}/\mu^{2}\right)  $ with $\mu\approx1$ GeV and $G^{2}%
D_{0}\approx30$ GeV$^{-1}\mu^{-1}$ is the non-perturbative gluon propagator.
The fact that we put inner quark lines on shell is expressed by $\delta\left(
q_{1}^{2}\right)  \delta\left(  q_{2}^{2}\right)  $.

After some approximations, based on the fact that integrand in
(\ref{Mfi3appen}) is exponentially damped in $\vec{v}^{2},$ what allows to
consider $\vec{v}$ as small, we obtain (for more details see \cite{Szere}):%

\[
M_{fi}^{\left(  3\right)  }=-A\frac{4}{12}\delta_{s,c}^{if}\delta_{uw}\int
d^{2}\vec{v}f\left(  \vec{v}\right)  \exp\left(  \frac{-\vec{v}^{2}}{\tau
}\right)  ,
\]
where:%

\begin{align*}
f\left(  \vec{v}\right)   &  =\left\{  p_{1}^{\beta}V_{\beta\xi\delta}\left(
l_{1},-r_{1},-q\right)  \frac{-g^{\delta\rho}}{q^{2}}V_{\rho\sigma\lambda
}\left(  q,-r_{2},l_{2}\right)  p_{2}^{\lambda}\varepsilon_{1}^{\xi
}\varepsilon_{2}^{\sigma}\right. \\
&  \left.  +p_{1}^{\beta}V_{\beta\xi\delta}\left(  l_{1},-r_{2},\tilde
{q}\right)  \frac{-g^{\delta\rho}}{\tilde{q}^{2}}V_{\rho\sigma\lambda}\left(
-\tilde{q},-r_{1},l_{2}\right)  p_{2}^{\lambda}\varepsilon_{1}^{\sigma
}\varepsilon_{2}^{\xi}\right\}  .
\end{align*}
Substituting (\ref{V}) and (\ref{parametr}) to the above formula, using
$p_{1}^{2}=p_{2}^{2}=0$ and $\varepsilon_{1}r_{1}=\varepsilon_{2}%
r_{2}=\varepsilon_{1}p_{1}=\varepsilon_{2}p_{1}=0$ what is a consequence of
(\ref{polariz}), we obtain (\ref{f(v)}).

Let us denote colour structure of the diagram represented by $M_{fi}^{\left(
4\right)  }$ as follows:

\begin{figure}[h]
\begin{center}
\includegraphics[width=3.90cm,height=3cm]{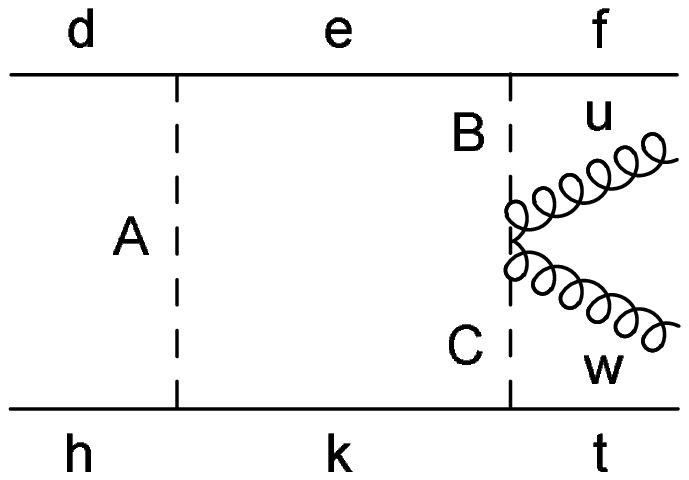}
\end{center}
\par
\vspace{0cm} \caption{The capital letters denote colour indicies of the
non-perturbative gluons and small letters colour indicies of quarks and
produced gluons.}%
\label{color3}%
\end{figure}

The matrix element $M_{fi}^{\left(  4\right)  }$ is given by (see Fig.
\ref{ok}):%

\begin{align}
M_{fi}^{\left(  4\right)  }  &  =-A\frac{1}{4}\left(  \lambda_{fe}^{B}%
\lambda_{ed}^{A}\right)  _{\text{singlet}}\frac{1}{4}\left(  \lambda_{tk}%
^{C}\lambda_{kh}^{A}\right)  _{\text{singlet}}%
{\displaystyle\int}
d^{4}Q\delta\left(  q_{1}^{2}\right)  \delta\left(  q_{2}^{2}\right)
\nonumber\\
&  D\left(  Q^{2}\right)  D\left(  l_{1}^{2}\right)  D\left(  l_{2}%
^{2}\right)  \bar{u}\left(  k_{1}\right)  \gamma^{\beta}\not q  _{1}%
\gamma^{\nu}u\left(  p_{1}\right)  \bar{u}\left(  k_{2}\right)  \gamma
^{\lambda}\not q  _{2}\gamma_{\nu}u\left(  p_{2}\right) \nonumber\\
&  \left\{  f^{xBC}f^{xuw}\left(  g_{\beta\xi}g_{\lambda\sigma}-g_{\beta
\sigma}g_{\xi\lambda}\right)  +f^{xBw}f^{xuC}\left(  g_{\beta\xi}%
g_{\lambda\sigma}-g_{\beta\lambda}g_{\xi\sigma}\right)  \right. \nonumber\\
&  +\left.  f^{xBu}f^{xCw}\left(  g_{\beta\lambda}g_{\xi\sigma}-g_{\beta
\sigma}g_{\xi\lambda}\right)  \right\}  \varepsilon_{1}^{\xi}\varepsilon
_{2}^{\sigma}.
\end{align}
Using (\ref{singlet}) and performing the rest colour calculations we obtain:%

\begin{align}
M_{fi}^{\left(  4\right)  }  &  =-\frac{A}{12}\delta_{c}^{if}\delta_{uw}%
{\displaystyle\int}
d^{4}Q\delta\left(  q_{1}^{2}\right)  \delta\left(  q_{2}^{2}\right)  D\left(
Q^{2}\right)  D\left(  l_{1}^{2}\right)  D\left(  l_{2}^{2}\right) \nonumber\\
&  \bar{u}\left(  k_{1}\right)  \gamma^{\beta}\not q  _{1}\gamma^{\nu}u\left(
p_{1}\right)  \bar{u}\left(  k_{2}\right)  \gamma^{\lambda}\not q  _{2}%
\gamma_{\nu}u\left(  p_{2}\right) \nonumber\\
&  \left[  2g_{\beta\lambda}g_{\xi\sigma}-g_{\beta\xi}g_{\lambda\sigma
}-g_{\beta\sigma}g_{\xi\lambda}\right]  \varepsilon_{1}^{\xi}\varepsilon
_{2}^{\sigma}.
\end{align}
After some approximations, fully described in \cite{Szere} we have:%

\[
M_{fi}^{\left(  4\right)  }=-A\frac{4}{12}\delta_{s,c}^{if}\delta
_{uw}s\varepsilon_{1}\varepsilon_{2}\int d^{2}\vec{v}\exp\left(
\frac{-\vec{v}^{2}}{\tau}\right)  ,
\]
what is our result (\ref{Mfi4-pocza}).

\end{document}